\documentclass[10pt,twocolumn,letterpaper]{article}

\usepackage{wacv}
\usepackage{times}
\usepackage{epsfig}
\usepackage{graphicx}
\usepackage{amsmath}
\usepackage{amssymb}
\usepackage{authblk}

\usepackage[sort&compress,numbers]{natbib}
\usepackage{subcaption}
\usepackage{graphicx}
\usepackage{amsmath}
\usepackage{amssymb}
\usepackage{booktabs}
\usepackage{dsfont}
\usepackage[misc,geometry]{ifsym} 
\usepackage{stmaryrd}
\usepackage{makecell}
\usepackage{multirow}
\usepackage{sg-macros}
\newcommand{\shortname}{{\it DeepCSR}}
\usepackage[misc]{ifsym}

\usepackage{xcolor}

\newcommand\extrafootertext[1]{%
    \bgroup
    \renewcommand\thefootnote{\fnsymbol{footnote}}%
    \renewcommand\thempfootnote{\fnsymbol{mpfootnote}}%
    \footnotetext[0]{#1}%
    \egroup
}
\def\wacvPaperID{71} 

\wacvfinalcopy 

\ifwacvfinal
\fi

\ifwacvfinal
\usepackage[breaklinks=true,bookmarks=false]{hyperref}
\else
\usepackage[pagebackref=true,breaklinks=true,colorlinks,bookmarks=false]{hyperref}
\fi

\begin{document}

\title{DeepCSR: A 3D Deep Learning Approach for Cortical Surface Reconstruction}

\author[1,2]{Rodrigo Santa Cruz}
\author[1,2]{Leo Lebrat}
\author[1]{Pierrick Bourgeat}
\author[2]{Clinton Fookes}
\author[1]{Jurgen Fripp}
\author[1,3]{Olivier Salvado}
\affil[1]{CSIRO Health \& Biosecurity, The Australian eHealth Research Centre, Brisbane, QLD, Australia}
\affil[2]{Image and Video Laboratory, Queensland University of Technology (QUT), Brisbane, QLD, Australia}
\affil[3]{CSIRO Data61, Brisbane, Australia}
\affil[ ]{\tt\small{rodrigo.santacruz@csiro.au}}

\maketitle
\thispagestyle{empty}

\begin{abstract}
The study of neurodegenerative diseases relies on the reconstruction and analysis of the brain cortex from magnetic resonance imaging (MRI). 
Traditional frameworks for this task like FreeSurfer demand lengthy runtimes, while its accelerated variant FastSurfer still relies on a voxel-wise segmentation which is limited by its resolution to capture narrow continuous objects as cortical surfaces.
Having these limitations in mind, we propose \shortname{}, a 3D deep learning framework for cortical surface reconstruction from MRI.
Towards this end, we train a neural network model with hypercolumn features to predict implicit surface representations for points in a brain template space. 
After training, the cortical surface at a desired level of detail is obtained by evaluating surface representations at specific coordinates, and subsequently applying a topology correction algorithm and an isosurface extraction method.
Thanks to the continuous nature of this approach and the efficacy of its hypercolumn features scheme, \shortname{} efficiently reconstructs cortical surfaces at high resolution capturing fine details in the cortical folding.
Moreover, \shortname{} is as accurate, more precise, and faster than the widely used FreeSurfer toolbox and its deep learning powered variant FastSurfer on reconstructing cortical surfaces from MRI which should facilitate large-scale medical studies and new healthcare applications.
\end{abstract}

\extrafootertext{Thanks to Maxwell plus:~\url{http://maxwellplus.com/}}

\section{Introduction}
\label{sec:intro}

\begin{figure*}[t!]
    \centering
    \includegraphics[width=\textwidth]{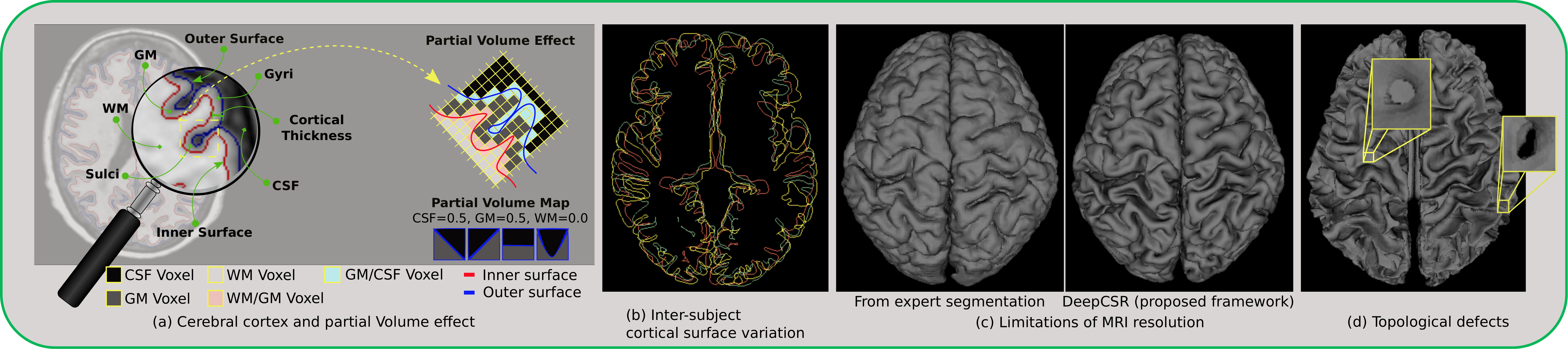}
    \vspace{-15px}
    \caption{Cerebral cortex and its reconstruction challenges. 
    (a) Depicts the brain cortex from an axial view. It also provides schematics for the partial volume effect (PVE) problem and the inaccurate boundary estimation using partial volume mapping. 
    (b) Overlays the cortical surfaces of co-registered MRIs from 3 different individuals exemplifying that the visual variability across individuals is like a fingerprint. 
    (c) Shows that meshes obtained from manual segmentation provided by experts, e.g. Neuromorphometrics, can not capture high-frequency details in the cortical surfaces. 
    (d) Present examples of small surface mesh defects that change the topology of the reconstructed surface.}
    \label{fig:intro_all}
\end{figure*}

The study of many neurodegenerative diseases and psychological disorders, rely on the analysis of the cerebral cortex using magnetic resonance imaging (MRI) \citep{Querbes:Brain2009,Dore:JAMA2013,Apostolova:JAMA2018}. 
As shown in \figref{fig:intro_all}a, the cortex can be visually described as the inner volume between two cortical surfaces: the inner surface which is the interface between the white matter (WM) and the gray matter (GM) tissues, and the outer surface which is the interface between the gray matter tissue and the cerebrospinal fluid (CSF).
Therefore, the goal of cortical surface reconstruction is to estimate reliable, accurate, and topologically correct~\citep{Pham:2010} triangular meshes, for the inner and outer cortical surfaces from a given MRI. 

Reconstructing cortical surfaces from MRI is a challenging problem.
First, cortical surfaces vary significantly across individuals as exemplified in \figref{fig:intro_all}b by overlaying axial slices of co-registered MRIs from 3 different patients. Second, voxels near the cortex tend to hold more than one tissue type due to the discrete nature of MRI and the folding patterns of the cortical surfaces which is known as partial volume effect (PVE) \citep{Ballester:MIL2002} in medical imaging. 
Consequently, sub-voxel variations of these surfaces can not be captured by pure voxel-wise segmentation approaches. This limitation leads to oversmoothed reconstructed cortical surfaces even when manual expert segmentation is employed as shown in \figref{fig:intro_all}c.
Finally, we need to enforce the spherical topology (\ie, homeomorphic to a sphere) of the reconstructed surfaces to allow surface-based analysis \citep{Schaer:MI2008,Rodriguez:NI2008} and visualizations \citep{Fischl:NI1999}. 
\figref{fig:intro_all}d emphasizes that arbitrarily small defects, known as \textit{holes}, can drastically change the topology of the reconstructed surface.

Despite these challenges, there exist frameworks for cortical reconstruction from MRI~\citep{Dale:NI1999,Macdonald:NI2000:ASP,Kriegeskorte:NI2001:BrainVoyage,Shattuck:MIA2002:BrainSuite,Kim:NI2005:CLASP,Dahnke:NI2013:CAT,Henschel:NI2020}. Traditionally, they consist of extensive pipelines of hand-crafted image processing algorithms subject to careful hyper-parameter tuning (\eg, thresholds, iteration numbers, and convergence criterion) and very long runtimes which prevent them from being employed in healthcare applications where immediate results are critical. For instance, the well-known FreeSurfer's V6.0~\citep{Dale:NI1999} takes around six hours per scan depending on the quality of the input MRI. 
Concurrently to our work, \citet{Henschel:NI2020} propose FastSurfer which alleviates this burden by integrating a fast and accurate deep learning based brain segmentation model to the FreeSurfer cortical surface reconstruction pipeline.  
However, these frameworks still rely on voxel-wise segmentation of the input MRI which has a limited resolution or employ partial volume mapping which also fails to define the tissue boundaries accurately. 
More specifically, multiple configurations of a tissue boundary have the same partial volume assignment as illustrated in \figref{fig:intro_all}a.

In this paper, we propose a 3D deep learning framework for cortical surface reconstruction from MR images named \textit{\shortname{}}. 
More specifically, we first reformulate this problem as the prediction of an implicit surface representation for points in a continuous coordinate system.
Then, the cortical surfaces are extracted using this implicit surface representation, a lightweight topological correction method, and an isosurface mesh extraction technique.
Since our framework predicts implicit surface representation for real-valued points instead of voxels, providing a continuous approximation to surfaces, we can reconstruct cortical surfaces at a higher resolution than the input MR image. This continuous formalism allows us to overcome the PVE problem without relying on inaccurate estimations like partial volume mapping.
We also develop a neural network architecture with hypercolumn features able to capture local information from the input MR image allowing the reconstruction of fine details describing the cortical folding.

In order to validate our approach, we first provide ablative studies of its main components. Then, we compare \shortname{} to the FreeSurfer V6.0 cross-sectional pipeline and FastSurfer for cortical reconstruction on the Test-Retest~\citep{Maclaren:TRT2014} and MALC~\citep{Landman:MICCAI2012,Marcus:2007} datasets.
We conclude that the proposed \shortname{} is able to reconstruct cortical surfaces faster and with higher reliability than the competitors, thus facilitating new healthcare applications. 


\section{Related Work}


\paragraph{Cortical surface reconstruction.} 
Traditionally, cortical surface reconstruction frameworks involve a lengthy sequence of image processing techniques~\citep{Dale:NI1999,Macdonald:NI2000:ASP,Kriegeskorte:NI2001:BrainVoyage,Shattuck:MIA2002:BrainSuite,Kim:NI2005:CLASP,Dahnke:NI2013:CAT}.
For instance, the widely used FreeSurfer \citep{Dale:NI1999} performs multiple image normalization routines, linear and non-linear registration techniques, brain tissue labeling, surface fitting models, and topology correction between other techniques to reconstruct the cortical surfaces. 
The other traditional frameworks follow a similar approach mostly differing in the mechanism used to fit the desired surfaces onto the segmented volumes. 
As examples, \citet{Kim:NI2005:CLASP} leverage a Laplacian field with stretch and self-proximity regularization terms to prevent mesh self-intersection and \citet{Han2004:NI2004:CRUISE} introduce a topology-preserving geometric deformable surface model. 
Therefore, these frameworks require a considerable processing time and eventually manual intervention of experts to fine-tune the parameters of some of these image processing techniques \citep{Iscan:2015} which limit their range of applications. 
Differently, we propose a swift framework able to reconstruct cortical surfaces efficiently by leveraging modern 3D deep learning models.

Concurrently to our work, \citet{Henschel:NI2020} propose a faster variant for the FreeSurfer pipeline named FastSurfer. 
In that work, expensive computations of FreeSurfer are replaced with modern and lightweight alternatives as a deep learning model for whole-brain segmentation and a spectral mesh processing algorithm for spherical mapping.
However, this approach still relies on voxel-wise brain segmentation which does not approximate continuous surfaces well due to its discrete nature as exemplified in \figref{fig:intro_all}c. In contrast, we propose a 3D deep learning model that directly predicts implicit surface representations for real coordinates in the brain MRI providing a continuous approximation to the cortical surfaces. Nonetheless, it is also important to acknowledge that FreeSurfer and FastSurfer are complete brain morphometry tools \citep{Spalletta:2018} providing further volumetric and surface-based analysis which is beyond the scope of this work. Note also that recent approaches tend to indicate that one could bypass the estimation of cortical surfaces directly predicting averaged morphometric measurements from the MR images~\citep{Rebsamen:Frontier2020,SantaCruz:Arxiv2020}.

\paragraph{Deep learning models for 3D reconstruction.} 
It comprises the set of deep learning models that learn  to reconstruct 3D geometries, \eg, objects and scenes, from data~\citep{Hoiem:IJCV2007,Saxena:PAMI2008}.
These deep learning models can be broadly categorized according to the 3D shape representation they operate on as either mesh-based, voxel-based, or implicit surfaces models.  
Mesh-based methods exploit explicit representations like template meshes \citep{Pontes:ACCV2018,Groueix:CVPR2018}, graphs \citep{Wang:ECCV2018, Wen:CVPR2019}, and parametric 3D models \citep{Kong:CVPR2017,Omran:3DV2018} to predict surfaces and object shapes. 
While these approaches are very convenient, they are also prone to produce noisy meshes with a large number of self-intersections which is very undesirable for cortical surface reconstruction. 
On the other hand, voxel-based methods reconstruct 3D shapes by predicting 3D voxel-grids of surface representations like discretized signed distances \citep{Park:CVPR2019} and level-sets \citep{Michalkiewicz:ICCV2019}. 
The main limitation of these approaches is to treat the output 3D space as a discretized grid that may not capture the fine details of the cortical surface folding. 
On the other hand, implicit surface models explore continuous representations of the 3D space where the objects are defined as 3D scalar fields \citep{Mescheder:CVPR2019,Park:CVPR2019}. 
These approaches are memory efficient and can explore the finer details of the target geometries. 
Therefore, we follow this approach and develop a novel deep learning model to reconstruct cortical surfaces from MR images.
\section{Method}

\begin{figure*}[t]
	\begin{center}		
		\includegraphics[width=\textwidth]{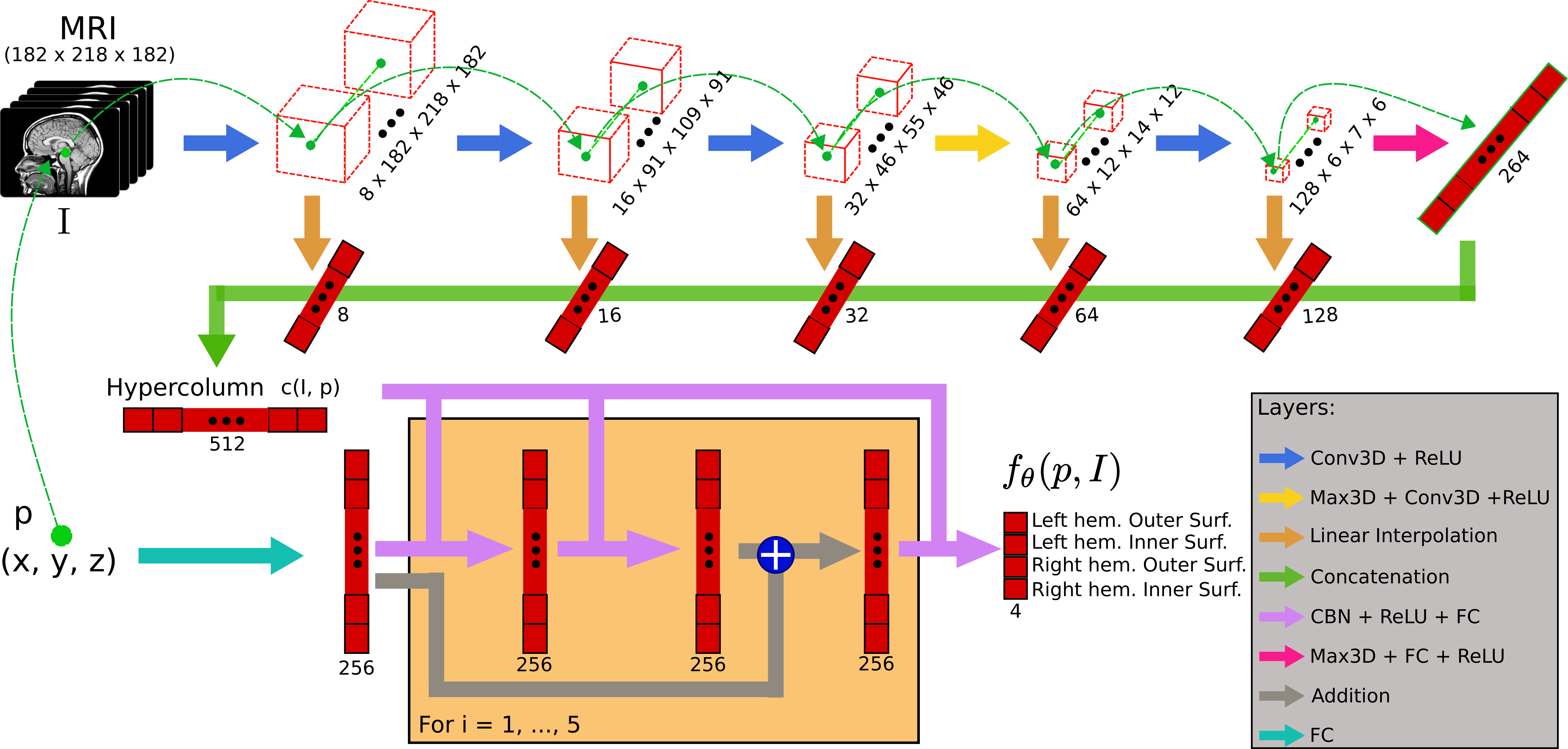}		
	\end{center}
 	\vspace{-10px}
	\caption{\shortname{}'s neural network architecture. It receives as input an image $I$ and the coordinates of a point $p$ in the template coordinate system $\Omega$. Then, it predicts the implicit surface representation $f_\theta(p, I)$ for the inner and outer cortical surfaces divided into left and right hemispheres.}
	\label{fig:method:model}
\end{figure*} 

In this section, we introduce \textit{\shortname{}}, a deep learning framework for cortical surface reconstruction from MR images.
First, we describe how to implicitly define surfaces by functions on the 3D Euclidean space, and how to learn these functions using a deep learning model. 
Then, we show how to perform cortical surface reconstruction using this model.

\subsection{Learning Implicit Surface Representation \label{sec:method:learning}}

In order to ease the presentation, we focus on the learning of a single target surface, but this framework can be generalized to multiple surfaces by introducing mild modifications.
Without loss of generality, one can implicitly define a Lipschitz surface $S \subset \reals^3$ by the $\ell$-level set of a continuous function $f_S: \reals^3 \mapsto \reals$ as,
\begin{equation}
    S = \left\lbrace p  \in \reals^3 \mid f_S(p) = \ell \right\rbrace,
    \label{eq_method:levelset}
\end{equation}
where the function $f_S$ maps to a scalar for every point $p \in \reals^3$ in the 3D Euclidean space. 
Furthermore, given $f_S$, the mesh representation of $S$ can be obtained by iso-surface extraction algorithms like marching cubes~\citep{Lewiner:2003} or ray casting~\citep{Wald:2005}. 
Therefore, reconstructing a specific surface can be thought as modeling a function such that its $\ell$-level set defines the desired geometry.

For a given surface $S$, the function $f_S$ can be modeled, for instance, using occupancy field or signed distance function.
The former is a binary valued function defined as, 
\begin{equation}
    r_S^{occ}(p) =  \ind{p \in S_{int}},
    \label{eq:method:occ}
\end{equation}
where $\ind{pred}$ is the indicator function evaluating to one, if its predicate $pred$ is true, and zero otherwise. In this case, the level set of interest to retrieve the surface is $\ell = \frac 12$. The occupancy field $r_S^{occ}$ divides the 3D space in the interior $S_{int}$ and exterior of the oriented surface $S$. On the other hand, the signed distance function consists of the Euclidean distance between every point in the 3D space and its projection, $\text{proj}_S(p) = \argmin_{q \in S} \norm{p-q}_2$, on the surface $S$.
Mathematically,
\begin{equation}
    r_S^{sdf}(p) = (2~r_S^{occ}(p)-1) \norm{p - \text{proj}_S(p)}_2.
    \label{eq:method:sdf}
\end{equation}
Therefore, ${r}_S^{sdf}$ assigns positive distances to points in the interior of the surface $S_{int}$, zero to points on the surface $S$ (\ie, $\ell = 0$), and negative distances otherwise. 
These tools are widely used in surface modeling, then we evaluate both in the context of cortical surface reconstruction.  

Our final goal is to reconstruct surfaces according to an observed MR image. 
Rather than conditioning the function $f_S$ on a specific surface, one should condition it on an input 3D MR image.
While this formulation seems logical, we notice that MR images may be defined in different coordinate systems producing an unbounded input space for learning. 
We overcome this difficulty by co-registering the input MR images to a brain template constraining the space where the target surfaces exists to the coordinates of the bounding box containing the template brain in the template coordinate system. 
Therefore, our problem boils down to model the function,
\begin{equation}
    f: \Omega \times \I \mapsto \reals,
    \label{eq:method:model}
\end{equation}
where $\Omega \subset \reals^3$ is a closed and bounded subspace where the template brain is defined and $\I$ denotes the space of MR images $I$ represented by a 3D grid of voxels.

Using a machine learning approach to model such a function, we first define a dataset $\D=(I_i, S_i)_{i=1, \ldots, N}$ of $N$ MR images $I_i$ and their corresponding surfaces $S_i$ which can be obtained using traditional pipelines for automatic cortical surface reconstruction~\citep{Dale:NI1999,Shattuck:MIA2002:BrainSuite,Dahnke:NI2013:CAT}. 
Then, we parametrize the function $f_\theta$ in terms of learnable parameters $\theta$ and minimize the regularized empirical risk on the dataset $\D$. This learning problem can be stated as,
\begin{equation}
\argmin_{\theta} ~ \frac{1}{N} \sum_{i=1}^{N}  \int_{\Omega} \Delta\left(r_{S_i}^*(p), f_\theta(p, I_i)\right) \text{d}p + \R(\theta),
\label{eq:method:objective}
\end{equation}
where $\R$ is some regularization function on the parameters $\theta$ and $\Delta$ is an appropriate cost function measuring discrepancies between the predicted implicit surface representation $f_\theta(p, I_i)$ and the true implicit surface representation $r_{S_i}^*(p)$ for the given point $p \in \Omega$, MR image $I_i$, and its corresponding surface $S_i$. In the case of occupancy field, we minimize the binary cross-entropy classification loss, while for signed distance function we opt for the $L^1$-loss.

In order to use this approach for cortical surface reconstruction, we extend the scalar-valued functions $f_\theta$ to a vector-valued function $f_\theta \in \reals^4$ where the predicted values are the surfaces representations for the inner and outer cortical surfaces further divided into left and right brain hemispheres. 
Consequently, the learning problem descried by~\eqnref{eq:method:objective} is solved jointly for these four cortical surfaces across all of the MRIs in the dataset $\D$. This formulation is an instance of multi-task learning which often produces models with good generalization ability~\citep{Ruder:2017}. 

In practice, we approximate the integral in \eqnref{eq:method:objective} by a finite sum of sampled locations $p$ in the reference space $\Omega$. 
We observed that the sampling scheme employed is critical to the quality of the reconstructed surface.
More specifically, as shown later in \secref{sec:exp:ablation}, the proposed framework reconstructs more accurate cortical surfaces when we sample more points near the surface, in addition to uniformly on the template space $\Omega$.
In order to sample points near the target surface, we first sample faces proportionally to their area, then we sample points uniformly in these faces using the triangle point picking method~\citep{Weisstein:1999}. 
Finally, we perturb these sampled points on the surface by adding a Gaussian $\mathcal{N}(0, 1)$ centered at zero with $1~mm$ variance getting points near the surface.
Such a sampling scheme provides global information to correctly align the predicted surface to the brain and a lot of local information to capture the highly folded geometry of the brain cortex.

\subsection{Neural Network Architecture and Hypercolumn Features \label{sec:method:model}}

We implement the parameterized function $f_{\theta}$ as an encoder-decoder neural network. 
The encoder takes as input a point $p$ represented by its coordinates in the MNI105 space and a co-registered MR image as a 3D voxel grid $I$. 
It first processes the input MR image by a sequence of 3D convolutional layers (Conv3D), ReLu activation functions, max pooling layers (Max3D), and fully connected layers (FC) producing multiple feature maps as indicated in the top of \figref{fig:method:model}. 
The convolutional layers use $3^3$ kernels, stride equal to two, and padding equal to one, with the exception of the first convolutional layer where the stride is equal to one. 
Our goal is to produce a rich hierarchy of visual features with different level of details.
Then, inspired by \citet{Hariharan:CVPR2015}, we project the input point coordinates into each feature map, linearly interpolate a feature value at these projected locations, and concatenate these interpolated values to form a 512-dimensional hypercolumn feature vector $c(I, p)$. 
This hypercolumn vector holds global and local visual cues for predicting the surface representation at the input point $p$ since it collects features from different levels of our hierarchy of features maps.
As shown later in \secref{sec:exp:ablation}, this architectural design is critical to capture the high frequency details on the target surfaces. 

On the other hand, the decoder receives as input the point coordinates $p$ and its corresponding hypercolumn vector $c(I, p)$ generated by the encoder for the image $I$ as previously described. 
The decoder processes these inputs through a sequence of fully connected layers, conditional batch normalization layers (CBN) \citep{Perez:2017} and ReLU activation functions as also indicated in \figref{fig:method:model}, outputting the predicted implicit surfaces representation $f_\theta(p, I) \in \reals^4$ for the inner and outer cortical surfaces divided into left and right brain hemispheres. 
The decoder network uses skip-connections and follows the method proposed by~\citet{Mescheder:CVPR2019}. 
However, our CBN layers compute a non-linear function conditioned on the hypercolumn vector $c(I, p)$ which encodes global and local visual cues for the prediction of the desired surfaces representation. 

We train such a model from scratch by optimizing the objective in \eqnref{eq:method:objective} extended for multiple surfaces using stochastic gradient descent. 
More specifically, we use Adam optimizer with an initial learning rate equal to $10^{-4}$ and back-propagate the loss computed on mini-batches of 5 MR images and 1024 sampled points per image from a precomputed pool of 4 million points whereby 10\% of them are sampled uniformly in $\Omega$ and 90\% are sampled near the target surfaces as explained in \secref{sec:method:learning}.

\subsection{Reconstructing Cortical Surfaces \label{sec:method:inferece}}

\begin{figure}[t]
	\begin{center}		
		\includegraphics[width=0.5\textwidth]{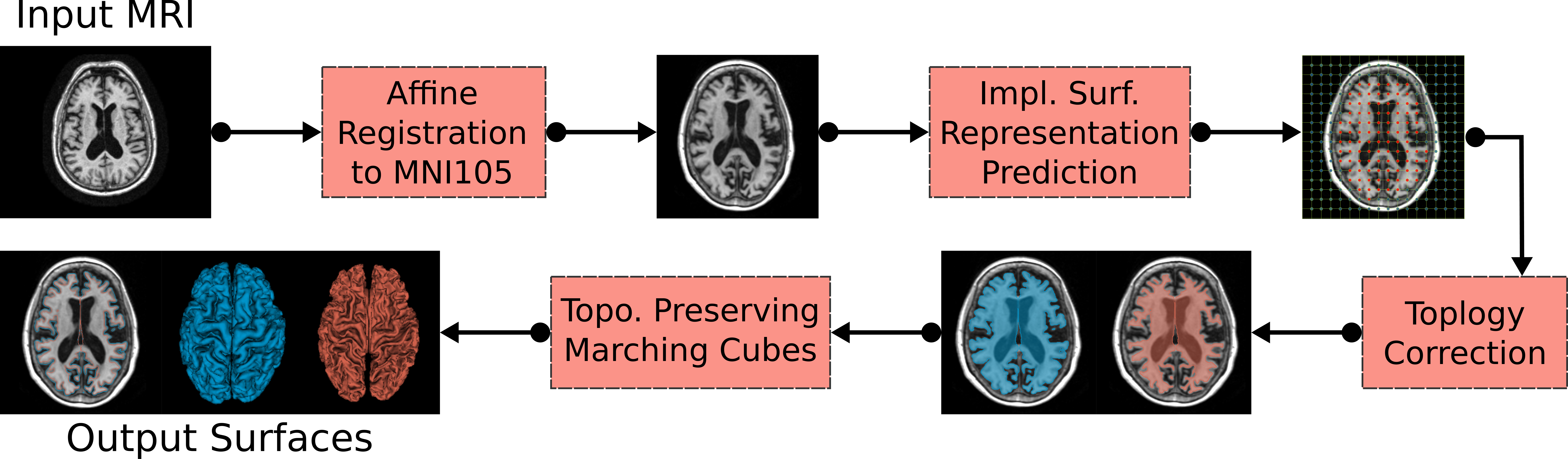}
	\end{center}
	\vspace{-15px}
	\caption{\shortname{} framework for cortical surface reconstruction from MRI.}
	\label{fig:method:overview}
\end{figure}

\shortname{} receives as input a MRI scan and outputs mesh representations for the outer and inner cortical surfaces further divided into the left and right brain hemispheres.
As illustrated in \figref{fig:method:overview}, we first perform an affine registration of the input MRI scan to the MNI105 brain template \citep{Mazziotta:NI1995}.
This registration aims to unify the coordinate systems across different MR scans easing the learning and prediction of implicit surfaces. 

Second, we construct a Cartesian grid of points at the desired resolution by dividing the template bounding box into evenly spaced points. 
Using the trained neural network $f_\theta$, we predict implicit surface representations at these points, represented by their continuous coordinates, for those four cortical surfaces according to the input MRI.
Throughout this paper we reconstruct cortical surfaces using $512^3$ points, unless mentioned, providing sub-voxel precision.
The model only needs to extract the feature maps once and can process the points in parallel, generating surfaces at high resolution efficiently. 

Third, since most of the applications of cortical surface reconstruction require surfaces with spherical topology (\ie, genus zero) as discussed in \secref{sec:intro}, we employ a lightweight topology correction algorithm to fix the topological defects caused by wrong predictions of the implicit surface representation.
Specifically, we use the method proposed by \citet{Bazin:2007} which consists of a spherical level-set evolution that avoids evolving over critical points \citep{Pham:2010}, outputting a 3D implicit surface volume with guaranteed spherical topology.
Each target surfaces' topology is corrected independently, and these tasks are performed in parallel for efficiency.
Note that the proposed framework is agnostic to the topology correction method allowing the application of other techniques.

\begin{table*}[t!]
\caption{Results of the ablation study on the proposed \shortname{} framework for cortical surface reconstruction from MRI using the ADNI study data~\citep{Jack2008:ADNI}.}
\resizebox{\textwidth}{!}{\begin{tabular}{l|ccc|ccc|ccc|ccc|}
                      & \multicolumn{3}{c|}{Left Outer Surface} & \multicolumn{3}{c|}{Right Outer Surface} & \multicolumn{3}{c|}{Left Inner Surface} & \multicolumn{3}{c|}{Right Inner Surface} \\
                      &  EMD & AD ($mm$) & HD ($mm$)  &  EMD & AD ($mm$) & HD ($mm$) &  EMD & AD ($mm$) & HD ($mm$) &  EMD & AD ($mm$) & HD ($mm$) \\
\hline

\shortname ~(SDF.)     &  \makecell{$\textbf{7.084}$ \\ $(\pm \textbf{0.797})$} &  \makecell{$\textbf{0.298}$ \\ $(\pm \textbf{0.184})$} &  \makecell{$\textbf{0.654}$ \\ $(\pm \textbf{0.596})$} & \makecell{$\textbf{7.083}$ \\ $(\pm \textbf{0.851})$} &  \makecell{$\textbf{0.294}$ \\ $(\pm \textbf{0.155})$} &  \makecell{$\textbf{0.651}$ \\ $(\pm \textbf{0.530})$} & \makecell{$\textbf{6.440}$ \\ $(\pm \textbf{0.810})$} &  \makecell{$\textbf{0.267}$ \\ $(\pm \textbf{0.216})$} &  \makecell{$\textbf{0.562}$ \\ $(\pm \textbf{0.725})$} & \makecell{$\textbf{6.401}$ \\ $(\pm \textbf{0.692})$} &  \makecell{$\textbf{0.260}$ \\ $(\pm \textbf{0.162})$} &  \makecell{$\textbf{0.542}$ \\ $(\pm \textbf{0.523})$} \\

\shortname ~(Occ.)     &  \makecell{$7.133$ \\ $(\pm 0.757)$} &  \makecell{$0.312$ \\ $(\pm 0.167)$} &  \makecell{$0.711$ \\ $(\pm 0.525)$} & \makecell{$7.108$ \\ $(\pm 0.857)$} &  \makecell{$0.305$ \\ $(\pm 0.134)$} &  \makecell{$0.672$ \\ $(\pm 0.554)$} & \makecell{$6.459$ \\ $(\pm 0.802)$} &  \makecell{$0.281$ \\ $(\pm 0.284)$} &  \makecell{$0.587$ \\ $(\pm 0.724)$} &  \makecell{$6.522$ \\ $(\pm 0.632)$} &  \makecell{$0.289$ \\ $(\pm 0.127)$} &  \makecell{$0.582$ \\ $(\pm 0.570)$}\\
\hline

Uniform Sampling      &  \makecell{$7.342$ \\ $(\pm 0.856)$} &  \makecell{$0.510$ \\ $(\pm 0.208)$} &  \makecell{$1.185$ \\ $(\pm 0.665)$} & \makecell{$7.415$ \\ $(\pm 0.868)$} &  \makecell{$0.538$ \\ $(\pm 0.169)$} &  \makecell{$1.326$ \\ $(\pm 0.524)$} & \makecell{$6.584$ \\ $(\pm 0.804)$} &  \makecell{$0.501$ \\ $(\pm 0.236)$} &  \makecell{$1.100$ \\ $(\pm 0.728)$} &   \makecell{$6.537$ \\ $(\pm 0.691)$} &  \makecell{$0.465$ \\ $(\pm 0.184)$} &  \makecell{$0.997$ \\ $(\pm 0.521)$} \\

No hypercolumns       &  \makecell{$10.892$ \\ $(\pm 1.375)$} &  \makecell{$5.145$ \\ $(\pm 0.489)$} &  \makecell{$12.252$ \\ $(\pm 1.267)$} & \makecell{$10.649$ \\ $(\pm 1.201)$} &  \makecell{$5.173$ \\ $(\pm 0.437)$} &  \makecell{$12.346$ \\ $(\pm 1.106)$} & \makecell{$15.644$ \\ $(\pm 1.914)$} &  \makecell{$5.629$ \\ $(\pm 0.603)$} &  \makecell{$12.811$ \\ $(\pm 1.254)$} & \makecell{$15.434$ \\ $(\pm 1.814)$} &  \makecell{$5.698$ \\ $(\pm 0.591)$} &  \makecell{$12.917$ \\ $(\pm 1.113)$}\\

Single Surface Model   & \makecell{$7.109$ \\ $(\pm 0.872)$} &  \makecell{$0.314$ \\ $(\pm 0.229)$} &  \makecell{$0.702$ \\ $(\pm 0.885)$} & \makecell{$7.088$ \\ $(\pm 1.042)$} &  \makecell{$0.323$ \\ $(\pm 0.249)$} &  \makecell{$0.731$ \\ $(\pm 0.958)$} & \makecell{$6.454$ \\ $(\pm 0.807)$} &  \makecell{$0.280$ \\ $(\pm 0.199)$} &  \makecell{$0.568$ \\ $(\pm 0.708)$} & \makecell{$6.429$ \\ $(\pm 0.709)$} &  \makecell{$0.278$ \\ $(\pm 0.186)$} &  \makecell{$0.564$ \\ $(\pm 0.606)$} \\
\hline
Resolution $64^3$ &  \makecell{$9.164$ \\ $(\pm 1.114)$} &  \makecell{$3.227$ \\ $(\pm 0.369)$} &  \makecell{$8.039$ \\ $(\pm 1.103)$} & \makecell{$8.992$ \\ $(\pm 1.068)$} &  \makecell{$3.268$ \\ $(\pm 0.344)$} &  \makecell{$8.123$ \\ $(\pm 1.047)$} &  \makecell{$8.318$ \\ $(\pm 0.962)$} &  \makecell{$2.889$ \\ $(\pm 0.365)$} &  \makecell{$6.189$ \\ $(\pm 0.950)$} &  \makecell{$8.299$ \\ $(\pm 0.893)$} &  \makecell{$2.996$ \\ $(\pm 0.325)$} &  \makecell{$6.429$ \\ $(\pm 0.828)$} \\
Resolution $128^3$ & \makecell{$8.340$ \\ $(\pm 1.117)$} &  \makecell{$1.400$ \\ $(\pm 0.314)$} &  \makecell{$3.879$ \\ $(\pm 1.023)$} & \makecell{$8.277$ \\ $(\pm 1.077)$} &  \makecell{$1.408$ \\ $(\pm 0.305)$} &  \makecell{$3.868$ \\ $(\pm 1.006)$} & \makecell{$6.680$ \\ $(\pm 0.776)$} &  \makecell{$0.872$ \\ $(\pm 0.261)$} &  \makecell{$1.929$ \\ $(\pm 0.779)$} &  \makecell{$6.681$ \\ $(\pm 0.669)$} &  \makecell{$0.893$ \\ $(\pm 0.207)$} &  \makecell{$1.982$ \\ $(\pm 0.618)$} \\
Resolution $256^3$ & \makecell{$7.349$ \\ $(\pm 0.882)$} &  \makecell{$0.453$ \\ $(\pm 0.198)$} &  \makecell{$1.112$ \\ $(\pm 0.649)$} & \makecell{$7.315$ \\ $(\pm 0.906)$} &  \makecell{$0.447$ \\ $(\pm 0.167)$} &  \makecell{$1.096$ \\ $(\pm 0.571)$} &  \makecell{$6.522$ \\ $(\pm 0.818)$} &  \makecell{$0.327$ \\ $(\pm 0.226)$} &  \makecell{$0.697$ \\ $(\pm 0.740)$} & \makecell{$6.525$ \\ $(\pm 0.652)$} &  \makecell{$0.324$ \\ $(\pm 0.168)$} &  \makecell{$0.691$ \\ $(\pm 0.549)$} \\
\hline
\end{tabular}}
\label{table:exp:ablation}
\end{table*}

Finally, in order to obtain the mesh representation of the target surfaces, we apply a topology preserving marching cubes algorithm \citep{Lewiner:2003}.
The predicted surfaces are in the template brain coordinate system, but these can be mapped back to the native space using the inverse of the transformation obtained during the registration of the input MR scan. 

\section{Experiments}
We now evaluate the performance of our method on the reconstruction of cortical surfaces from MR images.
In \secref{sec:exp:ablation}, we present an ablative study while, in \secref{sec:exp:freesurfer}, we compare it to the FreeSurfer V6 (cross-sectional pipeline) and FastSurfer. 

\subsection{Ablation Studies \label{sec:exp:ablation}}
\begin{figure*}[t!]
         \includegraphics[width=\textwidth]{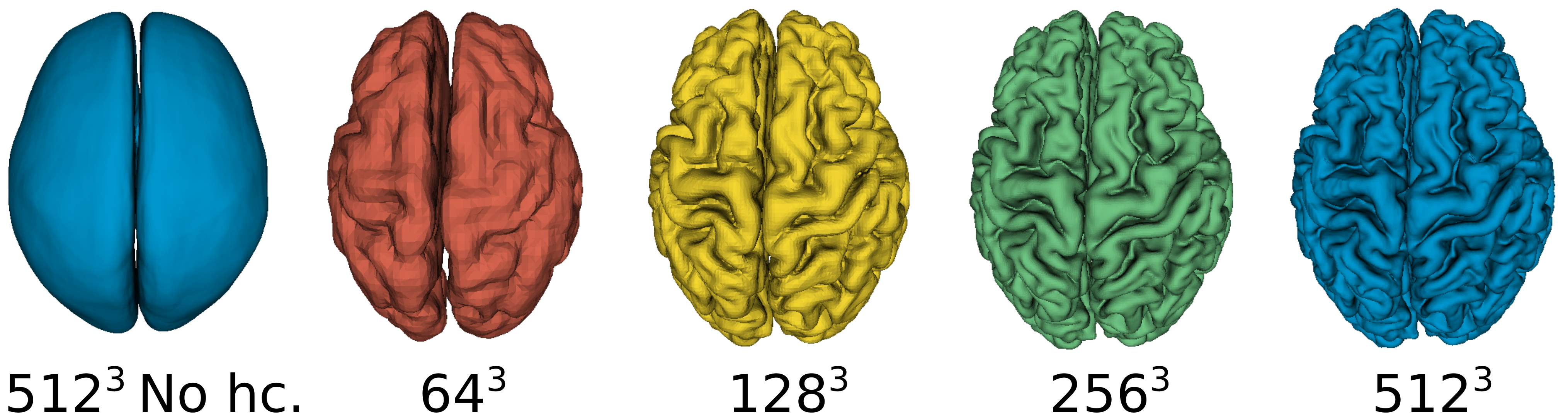} \\
         \vspace{-15px}
    	 \caption{Reconstructed outer cortical surfaces at different resolution or without hypercolumns ("No hc.").}
    	\label{fig:exp:ablation:qualitative}
\end{figure*}

We now perform experiments with variations of the proposed \shortname{} to measure the importance of its main components.
The quantitative results are presented in \tabref{table:exp:ablation}, while the qualitative results are shown in \figref{fig:exp:ablation:qualitative}. 
See below a brief summary of the dataset and evaluation metrics used, in addition to the detailed description of these experiments and the discussion of their results.

\noindent\textbf{Dataset.}
We use the MRI data provided by the Alzheimer's Disease Neuroimaging Initiative (ADNI)~\citep{Jack2008:ADNI} and their corresponding pseudo-ground truth surfaces generated with FreeSurfer V6.0.
The ADNI dataset consists of 3876 MRI images from 820 different subjects collected at different time points. 
We split this dataset by subjects obtaining 2353 MRI scans from 492 subjects for training ($\approx 60\%$), 375 MRI scans from 82 subjects for validation ($\approx 10\%$) and 1148 MRI scans from 246 subjects for testing ($\approx 30\%$). 
We train the models on the training set until their loss plateau on the validation set and report their performance on the test set.
We emphasize that these splits do not have MRIs or subjects in common for an unbiased evaluation.

\noindent\textbf{Evaluation metrics.}
We report the mean and standard deviation of well known surface comparison metrics like earth mover's distance (EMD), average absolute distance (AD), and Hausdorff distance (HD)~\citep{Taha:BMC2015,Rubner:1998,Tosun:NI2006}. 
EMD is the minimum amount of ``work'' to morph one surface to the other (lower is better), while AD and HD is the mean and maximum distance between closest points in two meshes (lower is better), respectively. Since HD is very sensitive to outliers, we use the $90^{th}$ percentile instead of the maximum as suggested by \citet{Huttenlocher:PAMI1993}.
It is also important to mention that AD and HD are computed in a bidirectional way for symmetry and over a sub-sample of 100k points.

\noindent\textbf{Signed distance vs. Occupancy Field.} 
We compare the performance of the \shortname{} framework when using signed distance function (first row in \tabref{table:exp:ablation}) and occupancy field (second row in \tabref{table:exp:ablation}) as implicit surface representation.
The signed distance function provides better results than occupancy field since such a representation offers more information than simple binary labels to the learning process and mesh extraction method.
More specifically, it also indicates the remoteness of the target surface at every point in the 3D space whereas the occupancy field just says whether a given point is inside or outside the target surface.

\noindent\textbf{Surface-based Sampling.} 
Despite uniform sampling has presented better results on generic object surface reconstruction as in \citet{Mescheder:CVPR2019}, it is not the best choice for cortical surfaces due to the high frequency details localized on the surface. 
To support such a claim, we train the \shortname{} framework to regress signed distance function of points sampled uniformly in the 3D template space and compare with the proposed sampling scheme described in \secref{sec:method:learning}. 
The results for the uniform sampling approach is reported in the third row of \tabref{table:exp:ablation}, while our surface-based sampling scheme is in the first row of the same table. 
The surface-based schema outperforms uniform sampling in all of the evaluation metrics. 

\noindent\textbf{Hypercolumn features.}
It is the most important component in our model, since we are not able to recover the high frequency details in the cortical surfaces without it. 
In order to support this claim, we remove the hypercolumn formation steps from our encoder depicted in \figref{fig:method:model}, reshape the encoder's last layer to output a 512-dimensional vector, use this vector as the decoder input, and train this model as before.
These changes remove the hypercolumn schema but keeps the model with similar capacity since no layer or weights are removed.
The performance of such a model is reported in the fourth row of \tabref{table:exp:ablation} and it is much worse than the proposed model for all the evaluation metrics. Furthermore, \figref{fig:exp:ablation:qualitative} depicts oversmoothed cortical surfaces that were reconstructed without hypercolumn features.

\noindent\textbf{Single Surface Model.} 
As previously observed by the machine learning community \citep{Ruder:2017}, multi-task learning tends to provide better generalization.
Similarly, we observe that learning the surface representations jointly produces better results than learn multiple models independently. 
We demonstrate such a result by training and evaluating a network for each target surface independently and comparing them to a network trained jointly for all the target surfaces. 
The results for the single-surface approach is presented in the fifth row of \tabref{table:exp:ablation}, while the multi-surface is presented in the first row of the same table. 

\noindent\textbf{Output Resolution.}
We reconstruct cortical surfaces at different resolutions by predicting implicit surface representations for 3D evenly spaced grids of points of different sizes as explained in \secref{sec:method:inferece}. 
\figref{fig:exp:ablation:qualitative} depicts cortical surfaces generated at $64^3$, $128^3$, $256^3$, and $512^3$ 3D evenly spaced grids of points.
As the resolution grows, our method reconstructs more details of the cortical surfaces.
The quantitative results are presented in the last three rows of \tabref{table:exp:ablation}.

\subsection{Comparison to FreeSurfer and FastSurfer \label{sec:exp:freesurfer}}

\begin{figure*}[t!]
	\begin{center}		
		\includegraphics[width=\textwidth]{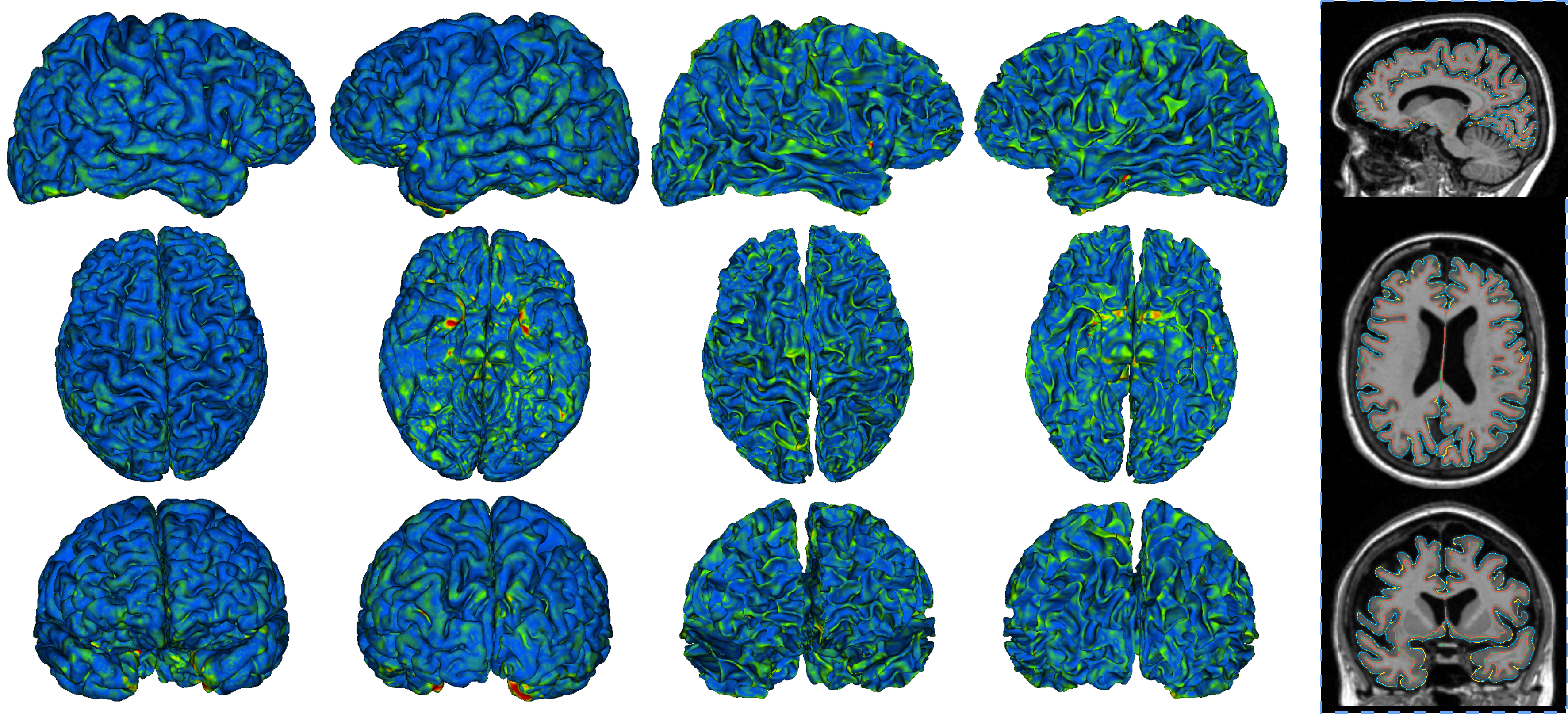}	\\
		\vspace{5px}
		\includegraphics[width=0.8\textwidth]{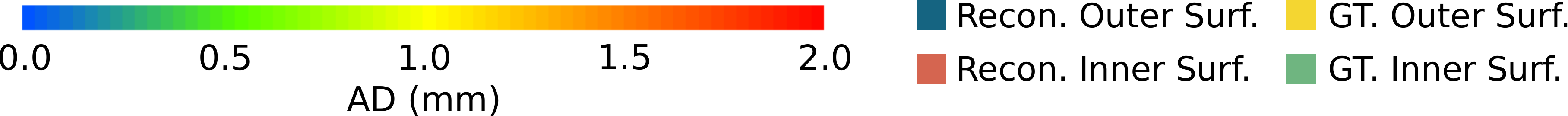} 	
	\end{center}
 	\vspace{-15px}
	\caption{Example of outer and inner cortical surfaces reconstructed with \shortname{}. 
	The surfaces are color coded with the absolute distance to the pseudo ground-truth surface.}
	\label{fig:exp:ablation:samples}
\end{figure*}

We now compare the precision, accuracy, and runtime of the proposed model to the FreeSurfer V6 (cross-sectional pipeline) and FastSurfer. 
More specifically, we use the \shortname{} (SDF) trained on the ADNI dataset and reconstruct surfaces at $512^3$ resolution as explained \secref{sec:method:inferece}.
\tabref{tab:freesurfer} summarizes the results and \figref{fig:exp:ablation:samples} shows examples of reconstructed surfaces.

\begin{table*}[t!]
\caption{Precision, accuracy and runtime comparison between FreeSurfer, FastSurfer, and \shortname{} on cortical reconstruction from MRI using Test-Retest (TRT) and  MALC datasets. 
}
\label{tab:freesurfer}
\centering
\resizebox{0.92\textwidth}{!}{\begin{tabular}{l|ccc|cc|c}
                  & \multicolumn{3}{c|}{\emph{Precision on TRT} } & \multicolumn{2}{c|}{\emph{Accuracy on MALC}} & \emph{Runtime}  \\
Method              & AD ($mm$) & $\% >1~ mm$ & $\%>2 mm$ & Dice & VS & $(minutes)$ \\ 
\hline
FreeSurfer    & \makecell{$0.241$ \\ $(\pm 0.291)$}& $2.472$ & $0.983$ & \makecell{$0.841$ \\ $(\pm 0.020)$} & \makecell{$0.953$ \\ $(\pm 0.027)$}  & \makecell{$373.86$ \\ $(\pm 47.64)$} \\
\hline

FastSurfer    & \makecell{$0.204$ \\ $(\pm 0.028)$}& $1.492$ & $0.374$ & \makecell{$0.834$ \\ $(\pm 0.021)$} & \makecell{$0.942$ \\ $(\pm 0.029)$}  & \makecell{$28.943$ \\ $(\pm 13.281)$} \\
\hline

\shortname{} & \makecell{$0.193$ \\ $(\pm 0.051)$} & $1.266$ & $0.263$ & \makecell{$0.846$ \\ $(\pm 0.019)$} & \makecell{$0.958$ \\ $(\pm 0.024)$} & \makecell{$27.824$ \\ $(\pm 1.393)$} \\
\hline
\end{tabular}}
\end{table*}

\noindent\textbf{Precision Analysis.} 
We measure the precision, \ie, repeatability, of the evaluated frameworks using the Test-Retest dataset (TRT) \citep{Maclaren:TRT2014} and the experimental protocol described by \citet{Tosun:NI2006}.
More specifically, the TRT dataset provides 120 T1-weighted MRI scans from 3 subjects which are scanned twice in 20 sessions spanning 31 days. 
Since no morphological changes should occur between successive scans of the same subject at the same session, the reconstructed surfaces should be identical up to the variations in the image acquisition and minor physiological changes.
Therefore, following the aforementioned protocol, we reconstruct the surfaces from every MRI scan using the evaluated frameworks, align pairs of surfaces from the same subject and session using the ICP algorithm, compute discrepancy measures between these pairs of aligned surfaces, and report group statistics of these metrics.
As discrepancy measures, we use the already discussed average absolute distance (AD), in millimeters ($mm$), between meshes and the percentage of distances greater than one (\% $>$ 1 $mm$) and two (\% $>$ 2 $mm$) millimeters as group statistics. 
The ``Precision'' column in \tabref{tab:freesurfer} summarizes the results for this experiment.
We observe that the proposed \shortname{} has better reproducibility than both FreeSurfer and FastSurfer which is critical for medical studies.

\noindent\textbf{Accuracy Analysis.} 
This experiment aims to measure how close the reconstructed cortical surfaces are to the true cortical surfaces. 
Since there is no manually annotated surface data for such a task, we decide to compare the evaluated methods on the segmentation of the brain cortex using the Multi-Atlas Labelling Challenge (MALC) dataset~\citep{Marcus:2007,Landman:MICCAI2012}. 
This dataset consists of 30 brain volumes \emph{manually segmented by experts} using the NeuroMorphometric labelling schema for the whole brain including cortical and subcortical structures. 
We first select all cortical labels to form the brain cortex ground-truth segmentation.
Then, for a given MR image, we reconstruct the inner and outer surfaces using the evaluated methods, discretize these surfaces to a grid of voxels of 1 $mm^3$, and resample these generated voxel-grids to the input MRI resolution and dimensions. 
Next, we remove the resampled voxel-grid of the inner surface from the interior of the resampled voxel-grid of the outer surface and perform morphological opening and dilatation to generate a surface-based brain cortex segmentation.
Finally, we report the Dice score (Dice) and volume similarity (VS) \citep{Taha:BMC2015} between the surface-based generated segmentation and the cortex ground-truth segmentation as an accuracy measure of the evaluated frameworks.
The ``Accuracy'' column in \tabref{tab:freesurfer} summarizes the results for this experiment. 
We observe that \shortname{} provides brain cortex segmentation with slightly greater overlap and more similar volume to the manually annotated data than the competitors.

\noindent\textbf{Runtime Analysis.} In order to compare the processing time required by the evaluated frameworks, we report the average elapsed time, in minutes, for these frameworks to reconstruct the cortical surfaces of the MRI scans in the MALC dataset~\citep{Marcus:2007,Landman:MICCAI2012}. 
The ``Runtime'' column in \tabref{tab:freesurfer} presents the results for this experiment. 
It also important to note that the FreeSufer and FastSurfer runtimes reported just takes into account the processing steps necessary to reconstruct the cortical surfaces and ignores any other computations for a fair comparison.
In summary, using a NVIDIA P100 GPU and Intel Xeon (E5-2690) CPU, our model is at least thirteen times faster than the FreeSurfer using the same hardware and input MRI scans.
When compared to the FastSurfer, the speed-up is much smaller but the variance in runtime is drastically reduced.
These improvements facilitate large medical studies and new healthcare applications. 

Moreover, profiling \shortname{}, we observe that the initial registration takes $2.334~ (\pm 0.022)$ minutes, the implicit surface prediction takes $8.672~(\pm 0.655)$ minutes, the topology correction takes $16.493~(\pm 1.064)$ minutes, and the marching cubes takes $0.325~(\pm 0.025)$ minutes. Therefore, further speed-up can be achieved by employing a faster topology correction method.
\section{Conclusion}

In this paper, we tackle the problem of directly reconstructing the brain cortex from MRI which is a critical task in clinical studies of neurodegenerative diseases.
We formulate this problem as the prediction of implicit surfaces for points in a continuous brain template coordinate system.
We also develop an encoder network architecture with hypercolumn features that is able to extract local and global image features from brain MR Images.
Due to the continuous nature of this formulation and the efficient design of our network, we are able to accurately reconstruct cortical surfaces at high resolution capturing the brain surface geometry.
Compared to the widely used FreeSurfer toolbox and its deep learning powered variant FastSurfer in two standard datasets, the proposed \shortname{} was found to be as accurate, more precise, and faster, which should facilitate large-scale medical studies and new healthcare applications.


{\small
\bibliographystyle{ieee_fullname}
\bibliography{long,biblio}
}

\end{document}